\documentstyle[12pt]{article}
\begin{document}
\begin{titlepage}
\vspace*{-1cm}
\begin{flushright}
DTP/96/76   \\
UR-1473 \\
ILL-(TH)-96-8\\
hep-ph/9609246 \\
August 1996 \\
\end{flushright}                                
\vskip 1.cm
\begin{center}                                                                  
{\Large\bf
Gluon Radiation in ${\rm t\bar{t}}$ Production
and Decay at the LHC}
\vskip 1.cm
{\large Lynne H. Orr}
\vskip .2cm
{\it Department of Physics, University of Rochester \\
Rochester, NY 14627-0171, USA }\\
\vskip   .4cm
{\large  T. Stelzer}
\vskip .2cm
{\it Department of Physics, University of Illinois-Urbana \\
Urbana, Illinois 61801, USA }\\
\vskip   .4cm
and
\vskip .4cm
{\large  W.J. Stirling}
\vskip .2cm
{\it Departments of Physics and Mathematical Sciences, University of Durham \\
Durham DH1 3LE, England }\\
\vskip 1cm                                                                    
\end{center}                                                                
\begin{abstract}
Understanding the pattern of gluon radiation in $t \bar t$ production
and decay processes is important for making an accurate determination
of the top mass from the momenta of its decay products.  The larger
energy of the LHC $pp$ collider boosts the top cross section by a
factor of 100 compared to that at the Tevatron, but it also increases
the amount of additional gluon radiation.  We calculate the cross
section for gluon radiation in top production and decay at the LHC.
The distributions of this radiation are presented and the exact
matrix-element results are compared with results from the HERWIG
parton-shower Monte Carlo.

\end{abstract}
\vfill
\end{titlepage}                                                                
\newpage                                                                       

\section{Introduction}
Reconstructing the top mass is critical to many of the physics goals
of the LHC.  The top-quark mass is interesting in its own right as a
fundamental parameter of the standard model (SM), and for its role in
pinning down other aspects of the SM and its extensions.  Our ability to
reconstruct $m_t$ also affects new physics search strategies, in which
top can appear as a signal or background.  Reconstructing the top mass
on an event by event basis is an important tool for distinguishing top
production from other processes.  Estimates for future runs at the
Tevatron, extrapolated from early Run 1 results,
suggest that  $m_t$ will be measured at the
4~GeV level \cite{TEV33}.  Recent improvements in Tevatron results 
\cite{WARSAW} may lead to more optimistic conclusions.
At the LHC, given the enhanced statistics, 
experiments may hope for an accuracy of 2~GeV \cite{mtLHC}.  
But the ability to do
this will depend on how well systematic effects --- especially those
associated with gluon radiation --- are understood and controlled.
It is therefore crucial to properly simulate the relevant physics.

By virtue of its energy and luminosity, the LHC will be a top factory.
The top cross section at the 14~TeV LHC is more than 100 times larger than
at the 2~TeV Tevatron.  This increase in production rate has a price, however: 
an increase in gluon radiation.  In a $t \bar t$ interaction, 7~TeV
protons can easily radiate quarks and gluons with only a
small penalty to be paid in the parton distributions.  As a result, at
the LHC there will be a plethora of radiation associated with top pair
production.  This radiation may well be the limiting factor in our
ability to reconstruct the top mass, both on an event by event basis,
and from global shapes.  For example, the sheer quantity of radiation
could result in emissions associated with top production (as opposed to decay)
 being included in the
$b$-quark jet cone, introducing spurious contributions to the top mass
reconstruction.  Additional jets may also introduce difficulties in
choosing the appropriate jets for reconstructing momenta in the
lepton+jets channel.  

This paper is an investigation of the effects of gluon radiation in
top events at the LHC.  In the next section, we compare $t\bar t$ and
$t\bar t g$ production at the LHC to that at the Tevatron.  We then
present in Section 3 a complete tree-level $\alpha_s^3$ calculation of
$p p \rightarrow W^+W^-b \bar b j$.  We present distributions of the
radiated jets, with the radiation decomposed into production- and
decay-stage emission.  In Section 4 we compare our matrix element
results with those from the parton shower Monte
Carlo HERWIG, and we comment on the 
precision to which HERWIG
appears to have this physics implemented for gluon radiation in
top-quark production and decay.  In Section 5 we present our
conclusions.

\goodbreak

\section{$t\bar t$ and $t \bar t g$ production at the Tevatron and LHC}

\begin{figure}
\vspace*{16cm}
\hspace*{-3.5cm}
\includegraphics{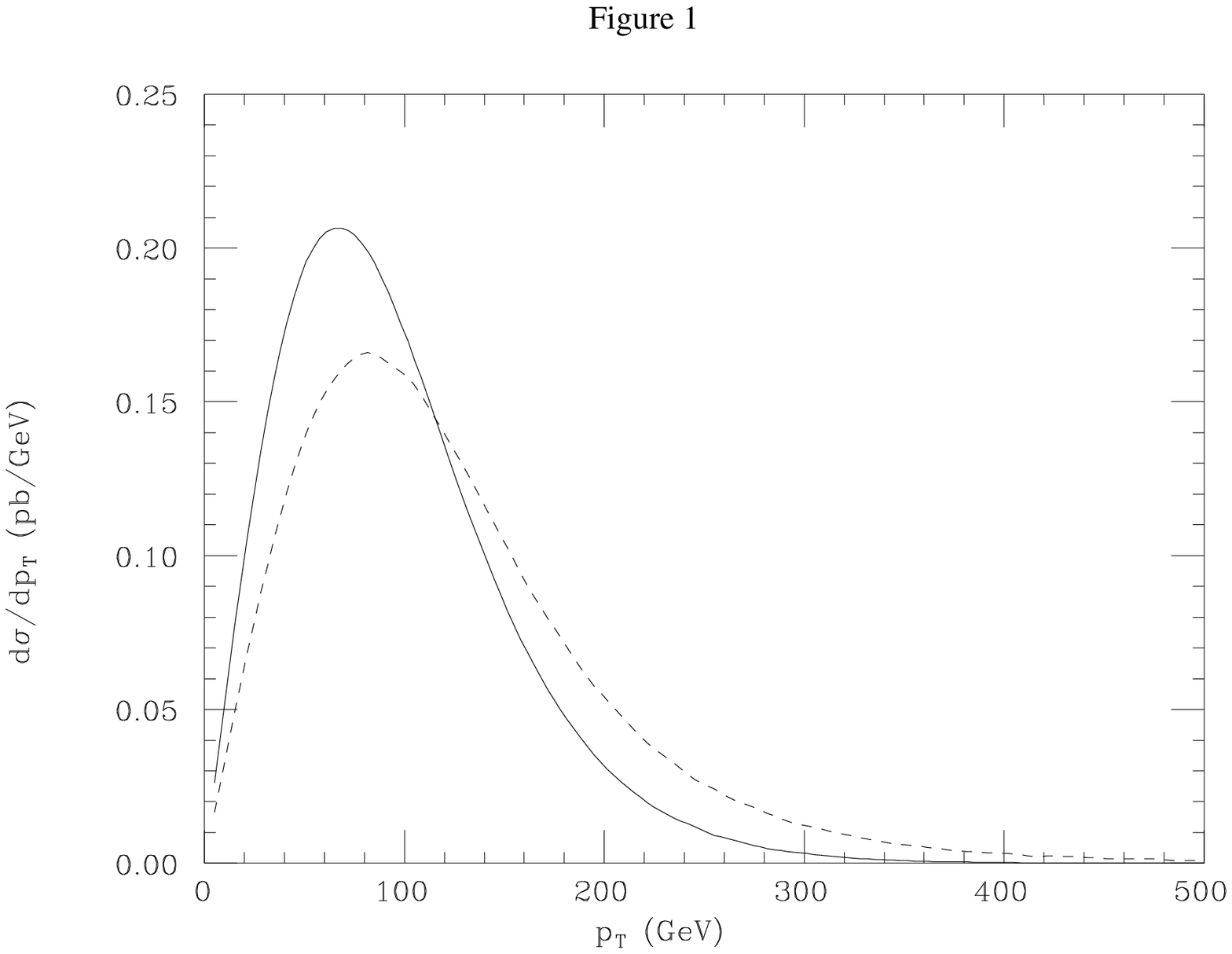}
\vspace{-7.5cm}
\caption{Distribution in top quark transverse momentum for $t\bar t$ production
at the Tevatron (solid line) and the LHC (dashed line).  The LHC curve
is normalized to the total $t \bar t$ cross section at the Tevatron.
\label{fig:pt}}
\end{figure}

To understand the effects of gluon radiation in top events at the LHC,
it is useful to compare top production there and at the Tevatron.  The
most obvious difference is in the $t \bar t$ production cross section,
which is on the order of 100 times higher at the LHC.  More relevant for our
purposes is a similarity between the two machines: the fact that for
heavy quark production, the mass of the quark rather than the collider
energy sets the scale for the quark's transverse momentum.  This is
illustrated in Figure~\ref{fig:pt}, which shows the transverse momentum of top
quarks produced at the Tevatron\footnote{With center-of-mass energy
1.8 TeV} (solid line) and LHC (dashed line),
normalized to the Tevatron cross section.  Despite the factor of seven
difference in collider energy, the transverse momentum of the top
quarks at the two machines is remarkably similar.  The only noticeable
effect of the LHC's higher energy is a slight spread in the
distribution at larger $p_T$.  (And although we do not show it here,
top quarks are produced at the LHC with broader rapidity distributions.)
Similar results are seen for 
$t \bar t j$ production.  

This similarity in top quark spectra at the two machines has several
consequences.  The most notable is the set of $x$ values at which the parton
distributions of the proton are probed at the two machines.  At the
Tevatron, the parton typically has a fraction  $x \simeq 0.2$ of the
proton's energy.  At the LHC, the typical value is only  $x \simeq
0.03$.  This results in 90\% of the top quarks produced at the
Tevatron coming from $q \bar q$ annihilation, whereas at the LHC about 90\%
of the top quarks come from the $gg$ initial state.  This means, among
other things, that we expect more gluon radiation in top production at
the LHC because of the gluons' larger color charge.  A second
consequence of the similarity in top quark spectra is that gluon
emission in top quark {\it decay} at the LHC should be similar to that
at the Tevatron.  At the LHC, therefore, gluon radiation is dominated
by production-stage emission.

\begin{figure}
\vspace*{16cm}
\hspace*{-3.5cm}
\includegraphics{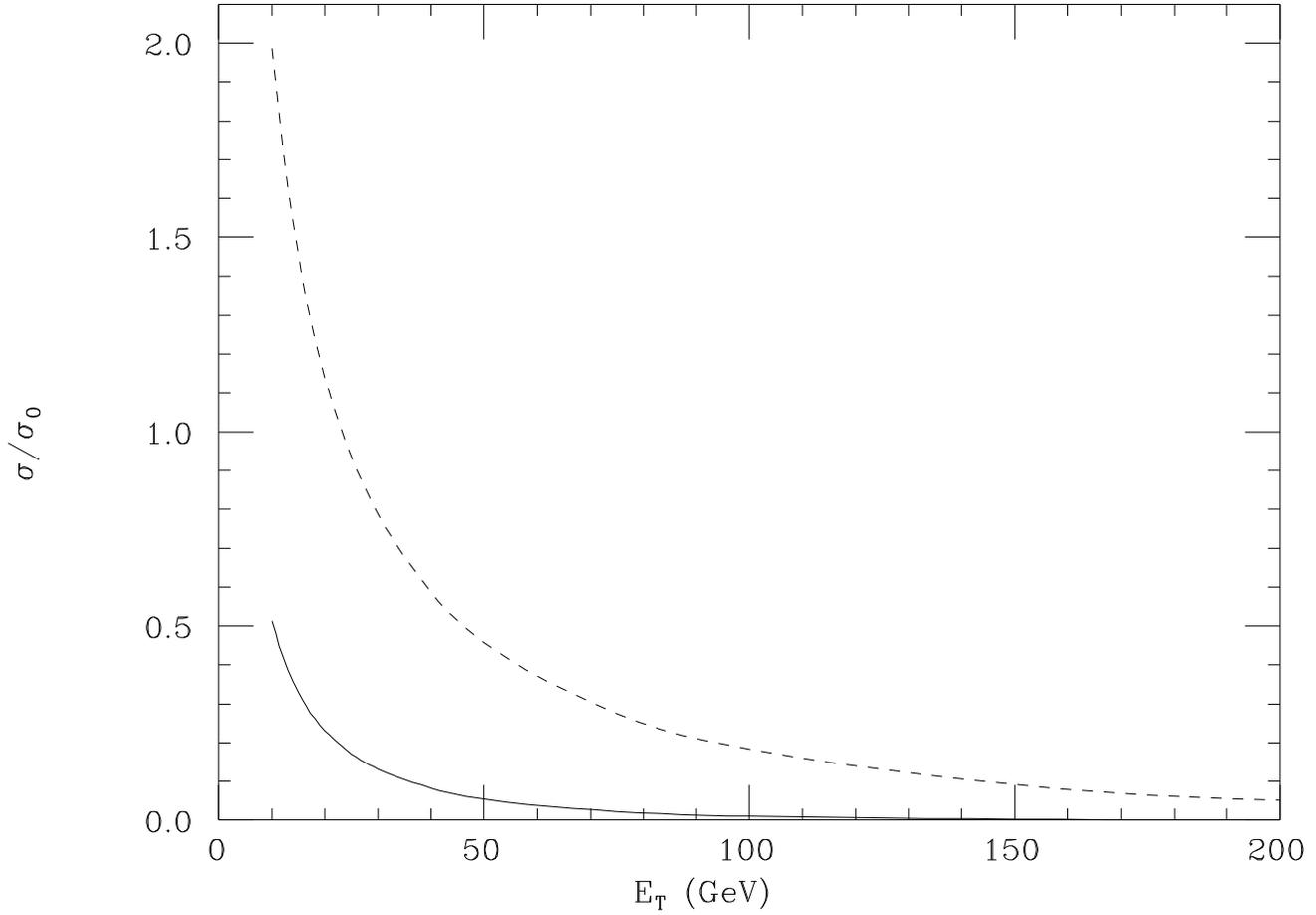}
\vspace{-7.5cm}
\caption{Ratio of $\sigma(t\bar t j$) to $\sigma(t \bar t)$ for $E_T^j >E_T$
at the Tevatron (solid line) and LHC (dashed line).  In both cases $t \bar t j$
production includes the subprocesses $q \bar q \to t \bar t g$,
$gg\to t \bar t g$, $qg\to t\bar t q$, and $\bar q g \to t\bar t \bar q$.
\label{fig:sigmas}}
\end{figure}

We will discuss the full gluon distributions below, but because of the
production-stage dominance we can draw a few general conclusions about
the importance of radiation in top events by considering $t \bar t j$
production.\footnote{In this calculation and in those below, we have chosen
the strong coupling constant scale as follows.    The factors of $\alpha_s$
associated with the lowest-order part of each process are evaluated
at $\sqrt{\hat{s}}$, the total subprocess center-of-mass energy.  The 
additional factor of $\alpha_s$ associated with emission of the extra jet is 
evaluated at the jet's transverse energy $E_{Tj}$.  Thus each 
$t\bar t j$ cross section contains an overall factor
$\alpha_s^2(\sqrt{\hat{s}}) \alpha_s(E_{Tj})$.}
Figure~\ref{fig:sigmas} shows the ratio of cross sections for $t \bar t
g$ and $t \bar t$ production at the Tevatron (solid line) and LHC
(dashed line) as a function of the minimum transverse energy 
$E_T$ of the gluon.  The
great enhancement for production stage emission at the LHC can be
attributed to two sources.  First, as mentioned above, gluons carry a
larger color charge than quarks.  Therefore the color in the $gg$
initial state at the LHC enhances gluon emission by a factor of
approximately $C_A/C_F= 9/4$ over the $q \bar q$ initial state at the
Tevatron.  Second, the parton distributions at the Tevatron fall very
steeply in the relevant $x$ range, making it difficult to
provide the additional energy required for a production stage
emission.  At the LHC, the additional energy can be obtained with less
of a cost in the parton densities.  

These effects are illustrated in the table, where we compare the 
contributions to $t \bar t$ and $t \bar t j$ production at the Tevatron
and LHC.  Cross sections are given in picobarns with cuts on the 
extra jet as indicated. 
If we compare the ratios $\sigma(t\bar t j)/\sigma(t \bar t)$ for the 
$q\bar q$ and $gg$ initial states, we see an enhancement for
$gg$ compared to $q \bar q$, as we would expect due to the larger color 
factor in the $gg$ initial state.  This happens for both colliders.
A closer look shows that the $gg$ 
enhancement is larger at the LHC, where we have, for $E_{Tj}>40\ {\rm GeV}$,
$\sigma(gg\to t\bar t g)/\sigma(gg\to t\bar t) = 0.46$ and
$\sigma(q\bar q\to t\bar t g)/\sigma(q\bar q\to t\bar t)=0.16$.  
At the Tevatron, these ratios are, respectively, 0.1 and 0.07.
The larger increase in the $gg$ cross section over $q \bar q$ at the LHC is
due to the diffference in behavior of the parton distributions discussed 
above.

\begin{center}
\begin{tabular}{|c|c|c|c|c|}  \hline                                    
\rule[-1.2ex]{0mm}{4ex}  
 & & $q\bar q$ & $gg$ & $qg, \bar q g$ \\ \hline 
 Tevatron & $t \bar t$ & 2.4     &  0.2     & -   \\
 & $t \bar t j, E_{Tj}> 10\ {\rm GeV}$ & 1.1   & 0.2  & 0.1     \\
 & $t \bar t j, E_{Tj}> 40\ {\rm GeV}$ & 0.17   & 0.02  & 0.03     \\ \hline
 LHC & $t \bar t$ &  50    &   330    &  -  \\
 & $t \bar t j, E_{Tj}> 10\ {\rm GeV}$ & 35  & 590 & 146      \\
 & $t \bar t j, E_{Tj}> 40\ {\rm GeV}$ & 8   & 151   & 63     \\
  \hline                                                                        
\end{tabular}
\end{center}

Although Fig.~\ref{fig:sigmas} provides an indication of the relative 
importance of
gluon radiation at the Tevatron and LHC, it it should not be taken too
literally.  For example, it should not be translated directly into an
expected number of top events containing an extra gluon.  There are
several reasons for this.  First, only production-stage radiation is
explicitly included.  Second, and more important, it represents a
fixed-order matrix element calculation which includes neither virtual
effects nor effects due to multiple gluon emission, both of which can
be important for low gluon energies.

In fact the figure serves as a guide to the regions where we can and
cannot trust the matrix-element results.  Roughly speaking, they are
reliable when the $t \bar t g$ cross section is well below $\sigma(t
\bar t)$.  This is satisfied at the Tevatron for all $E_T$ cuts shown.
At the LHC, however, the first-order cross section rises dramatically
with decreasing $E_T$ cut, and the $t \bar t g$ cross section with
gluon transverse energies greater than 10 GeV {\it exceeds} the
lowest-order cross section by a factor of 2, as can be seen in 
Fig.~\ref{fig:sigmas} and the table.
Clearly virtual and multi-gluon effects
must be important there.  We therefore restrict our LHC analysis to
gluons with transverse momentum greater than 40 GeV in what follows.

\section{Gluon radiation in top production and decay}

It is useful to distinguish between two different types of radiation
in $ t \bar t$ processes, as we have implicitly done above
and as has been discussed in previous work \cite{KOSetc,MOS,OSS}.
Gluons can be radiated in either the top
production or decay stages.
Production-stage emission occurs before the
top quark goes on shell and decay-stage emission occurs only after
the top quark goes on shell.  In principle, an event with an extra jet
can be classified as `production' or 
`decay' by looking at the invariant mass of the decay products.  In
production emission events, the $W$ and $b$ momenta will combine to give
the top momenta. In decay emission events, the gluon momentum must also be
included to reconstruct the top momenta.  

This interpretation is exact at the parton level in the narrow width
approximation.  Finite top width effects can blur this interpretation due to
interferences between production-and decay-stage emissions.  However,
the classification is still useful in our case because the top width
of 1.5~GeV is small compared to the 40~GeV gluon $E_T$ cut imposed in
the matrix element calculations.  It should be kept in mind that this
applies at the level of theory.  In an experiment, the
production-decay distinction is further blurred by jet energy
resolution and ambiguities associated with combinatorics and the like.

We have performed a complete tree-level $O(\alpha_s^3)$ calculation of
$p p \rightarrow W^+W^-b \bar b j$ at 14 TeV collision energy.  The
calculation was performed as in \cite{OSS}, with the exception of the 
choice of $\alpha_s$ scale as discussed above.  We  include all contributing
diagrams\footnote{We include all processes that give rise to an extra
jet: $q\bar{q} \rightarrow b\bar{b} W^+W^- g,$ $gg \to b\bar{b} W^+W^-
g,$ and $q g (\bar{q} g) \to b\bar{b} W^+W^- q (\bar q)$.} 
and their interferences (with helicity
amplitudes generated by MadGraph \cite{MADGRAPH}), and all top width
and $b$ mass effects.  Note that we do {\it not} include radiation off
the $W$ decay products.  We use MRS(A$'$) parton distributions
\cite{MRSA}.  The kinematic cuts imposed on the final-state partons
are (the subscript $j$ refers to the extra jet only):\footnote{The
cuts are applied to both the $b$ and $\bar b$ quarks.}
\begin{eqnarray}
 |\eta_j| \> & \leq & \> 3 \; ,\nonumber \\
 |\eta_b| \> & \leq & \> 2 \; ,\nonumber \\
E_{Tj}, \> & \geq & \> 40 \ {\rm GeV} \; , \nonumber \\
E_{Tb}, \> & \geq & \> 20 \ {\rm GeV} \; , \nonumber \\
\Delta R_{bj}, \Delta R_{b\bar b}  \> &  \geq  & \> 0.4  \; .
\label{cuts}
\end{eqnarray}

\begin{figure}
\vspace*{20cm}
\hspace*{-3.cm}
\includegraphics{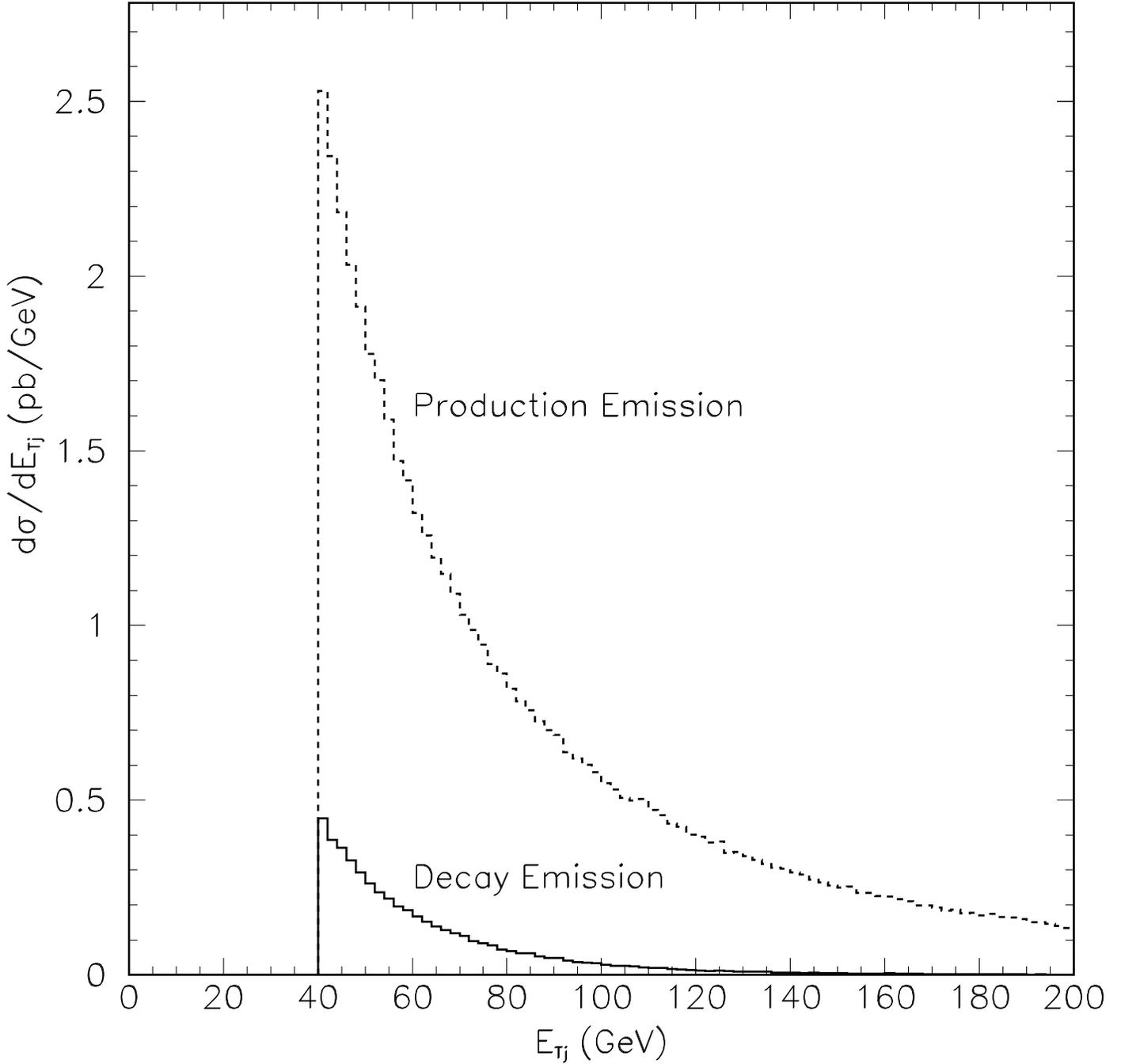}
\vspace{-4cm}
\caption[f3]
{Transverse energy distribution for jets produced in association with 
top production and decay,
via the subprocesses
$q \bar q, gg \to b W^+ \bar b W^- g$ and 
$qg (\bar q g) \to b W^+ \bar b W^- q (\bar q)$, 
in $pp$ collisions at $\sqrt{s} =14\ {\rm TeV}$.  
Contributions from production- (dashed histogram),
and decay-stage (solid histogram) emissions are shown.
The cuts are listed in  Eq.~(\ref{cuts}).
\label{fig:et}}
\end{figure}

\begin{figure}
\vspace*{20cm}
\hspace*{-3.cm}
\includegraphics{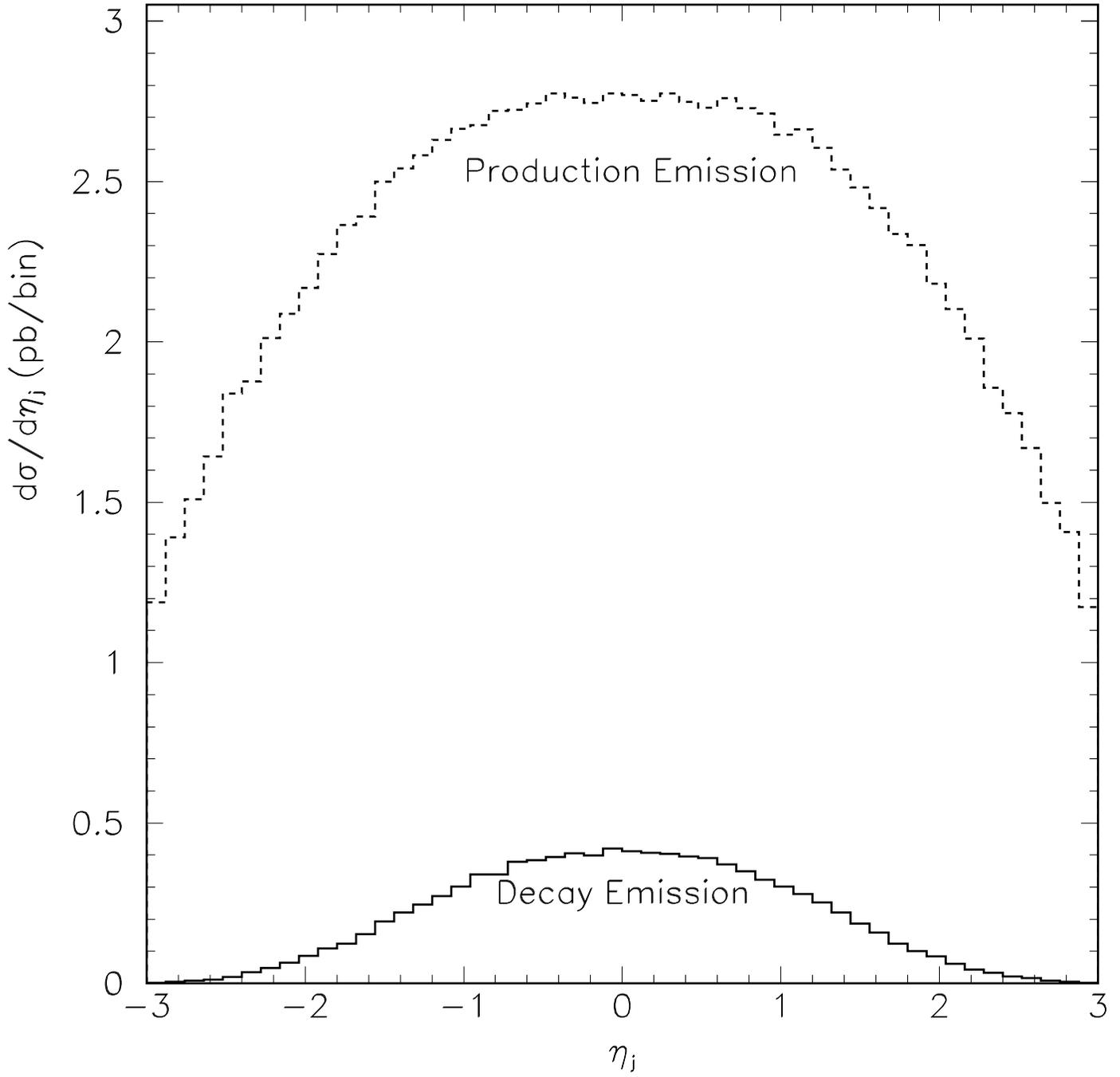}
\vspace{-4cm}
\caption[f4]
{Pseudorapidity distribution for jets produced in association with
top production and decay at the LHC.
Contributions from production- (dashed histogram),
and decay-stage (solid histogram) emissions are shown.
The cuts are listed in  Eq.~(\ref{cuts}).
\label{fig:eta}}
\end{figure}

\begin{figure}
\vspace*{20cm}
\hspace*{-3.cm}
\includegraphics{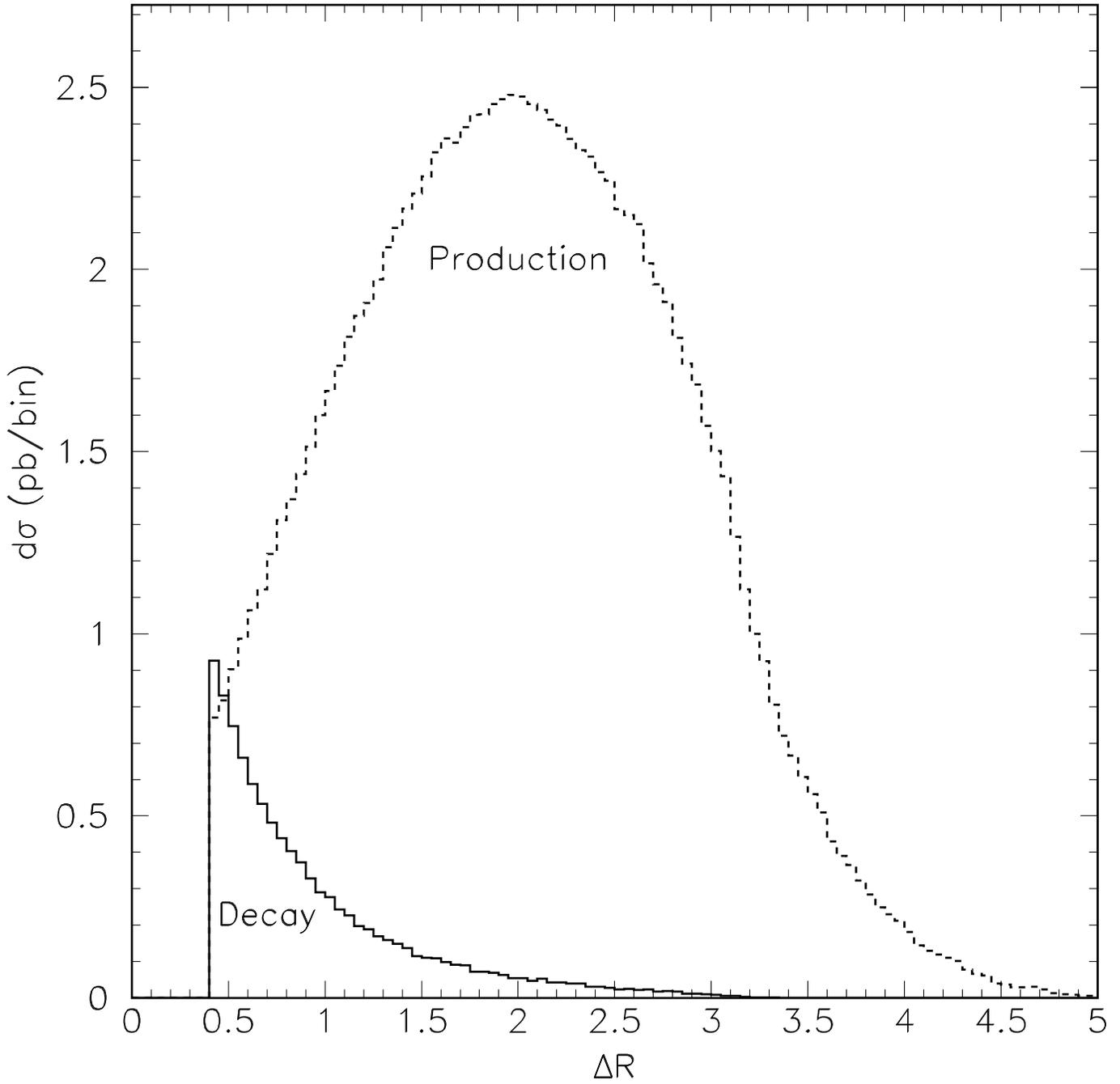}
\vspace{-4cm}
\caption[f5]
{Distribution in the jet-$b$ angular separation ($\Delta R_{bj}
 = (\Delta\eta_{bj}^2 + \Delta\phi_{bj}^2)^{1/2}$)
for jets produced in association with
top production and decay at the LHC.
Contributions from production- (dashed histogram),
and decay-stage (solid histogram) emissions are shown.
The cuts are listed in  Eq.~(\ref{cuts}).
\label{fig:dr}}
\end{figure}

The resulting distributions for the extra jet at the LHC are shown in 
Figures \ref{fig:et}--\ref{fig:dr}.
In each figure the distribution is decomposed into contributions from
production- (dashed line) and decay-stage (solid line) radiation
according to final-state kinematics
as described in \cite{OSS}.  
The most obvious feature of these distributions is the dominance of production
over decay emission, due to the enhancements in production
emission discussed above. 
The decay contribution does not receive this enhancement because its
behavior is determined not by the collider energy, but by the
phase space of a 175~GeV top-quark decay. 

In addition to the relative size, the kinematics of the two types
of emission are also interesting.  Figure \ref{fig:et} 
shows the jet $E_T$ distribution.
Both contributions fall off with increasing $E_T$, but production emission
extends to much higher values.  The smaller values of $E_T$ to which decay 
emission is constrained are again the consequence of the top 
decay kinematics.  Recall that even at the LHC, top quarks are produced with 
relatively modest transverse momentum (cf. Fig.~\ref{fig:pt}), 
so that gluons from the 
decay do not receive much of a boost in $E_t$.
Note also that an increase in the $E_T$ cut on the jet would result in 
a further reduction in relative size of the decay contribution compared
to production.

Figure \ref{fig:eta} shows the distribution in pseudorapidity of the extra jet.
Production emission is relatively flat in rapidity, as compared to the
more central decay emission.  This is consistent with our basic intuition
that decay-stage radiation, being associated with the final-state particles
--- which tend to appear in the central rapidity region --- is
also likely to be produced centrally.  But this decay contribution is
small; the important point to note here is that even in the central region,
it is production-stage radiation that dominates.

The tendency of decay-stage radiation to be associated with the final-state
$b$ quarks might lead one to expect that 
if the extra jet is `near' the $b$ jet it should be
included in the mass reconstruction, and if it is not it should be
excluded.  Figure \ref{fig:dr}, which shows the distribution in 
$\Delta R$ between the jet and the nearest $b$ quark,
confirms that the decay-stage radiation peaks close to the $b$ and 
production-stage radiation peaks further away.  Unfortunately, 
the production contribution is so large that it dominates even at 
the low $\Delta R$ cutoff.  A higher $E_T$ cut on the jet would make this situation
even worse.  The best 
choice of what is `near' the $b$ quark will therefore balance the competing
effects of decay emission falling outside the cone, and production
emission falling inside the cone.

It is tempting at this point to provide a prescription for 
dealing with the extra jets expected in top events at the LHC, for
example by specifying
how to make the best choice of what is `near' the $b$ quark.
But optimizing this choice at the parton level would be naive, because 
effects of multiple emissions, hadronization, and detector resolution
will all affect the results.

We also note that radiation from $W$ decay products has not been included in 
our analysis here.  Since the best top mass reconstruction is obtained 
in the lepton+jets mode, radiation from hadronic $W$ decays must
ultimately be included.  This  calculation has been done in the soft
gluon approximation \cite{MOS}, and the contribution from a single
hadronically decaying  $W$ is found to be substantial --- comparable in 
size and shape to the {\it total}  decay contributions from radiation off
the $tb$ and $\bar t \bar b$ antennae.  The exact calculation 
including hadronic $W$ decays is currently in progress \cite{OSSINPROG}.

In practice the effects of gluon radiation are incorporated into
the predictions that are used in experimental fits.  
The parton level calculation can and should be used
to ensure that the radiation physics is properly implemented in event
generators used in the experimental analysis. 

\section{Comparison with HERWIG}

Because the experimental analysis must rely on the 
predictions of Monte Carlo programs --- for example, in fits to three-jet
invariant mass distributions for top mass determination --- it is 
important that these programs contain the correct physics.  
The Monte Carlo program HERWIG \cite{HERWIG}, which is widely used in
experimental analyses, treats gluon radiation in top production and decay
using parton showers.  

In previous work \cite{OSS,OSSBIS}, we compared our results for
radiation at the Tevatron with predictions of version 5.8 of HERWIG.
We found significant discrepancies in regions where the two should
agree.  HERWIG appeared to have a deficit in decay-stage radiation
compared to production \cite{OSS}.  Further investigation revealed
differences even for $t \bar t g$ production at $e^+e^-$ colliders
\cite{OSSBIS}.  Recently HERWIG 5.8 was found to contain a bug
\cite{BUG,HERWIGNEW} which resulted in suppression of decay-stage radiation in
top events.\footnote{The bug appeared only in version 5.8; it was not
present in HERWIG 5.7 \cite{HERWIGNEW}.}  Here we continue our comparison of
matrix-element and parton-shower results using 
HERWIG 5.9 \cite{HERWIGNEW}, in which the bug has been corrected.  
Although we see some improvement in the agreement,
major differences still exist.

We begin by reproducing the LHC jet distributions 
using HERWIG 5.9. The details of comparing
a full parton shower Monte Carlo with a fixed-order matrix element
calculation were discussed in previous papers \cite{OSS,OSSBIS}.  The
idea is to combine particles from the parton shower into jets, and
compare distributions of these jets to those from the matrix element
calculation.  For hadron colliders we use a cone algorithm to combine
partons into jets.  We identify events with production- and
decay-stage emission according to the final state kinematics, as
described above.

\begin{figure}
\vspace*{20cm}
\hspace*{-3.cm}
\includegraphics{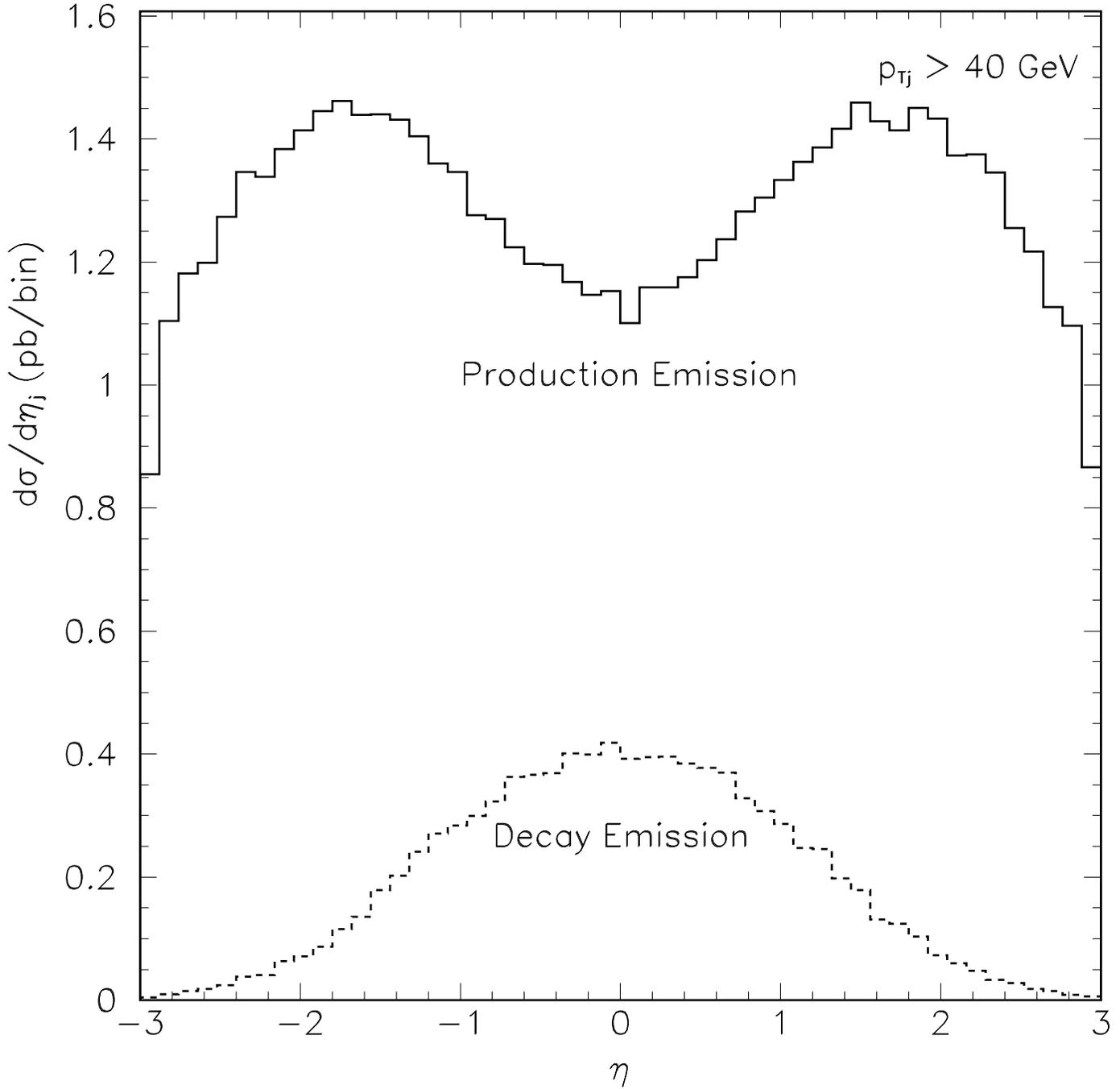}
\vspace{-4cm}
\caption[f6]
{Jet pseudorapidity distribution for extra jets in top production and 
decay at the LHC  as obtained 
using the HERWIG parton-shower Monte Carlo program,
version 5.9.  
Contributions from production- (solid histogram),
and decay-stage (dashed histogram) emissions are shown.
The cuts are listed in  Eq.~(\ref{cuts}).
\label{fig:etahw}}
\end{figure}

Results for the jet pseudorapidity are shown in Figure \ref{fig:etahw}.  
There is general agreement
between the matrix element calculation and the parton shower.
However, a closer examination reveals two important differences.
Looking at production
emission, we see that the HERWIG distribution is peaked at large 
$|\eta|$ and has a dip in the center.  In contrast, the matrix element 
distribution is relatively flat, as we have seen in Fig.~\ref{fig:eta}.  
Since the jets in this case have relatively
strong cuts, the perturbation series should be converging quickly and
the tree-level matrix element distributions should be accurate.  
This suggests that the approximations used in HERWIG may be responsible 
for this discrepancy.  

The second difference between the matrix-element and parton-shower results
is, as before, in the relative amounts of production  and decay
emissions.  Whereas HERWIG 5.8 with the bug had too {\it little}
decay-stage radiation, the corrected version now seems to have 
too {\it much} compared to the matrix element calculation.

This effect is illustrated more clearly in a simpler example.  As in
our previous work \cite{OSSBIS}, we simplify the comparison by looking
at $e^+e^-$ machines near $t \bar t$ threshold.  While the parton
calculation is an inclusive calculation with a fixed number of final
state particles, HERWIG is an exclusive calculation with an arbitrary
number of final particles.  To perform a meaningful comparison we
employ the Durham ($k_T$) successive recombination algorithm to reconstruct
jets from the HERWIG output \cite{DURHAM}.  In addition, we impose
cuts ($E_{Tj} > 10$~GeV, $\Delta R > 0.4$) on the jets to ensure the
matrix element is being evaluated in a region where the perturbation
series converges rapidly.  The validity of this comparison was
demonstrated in \cite{OSSBIS}.

\begin{figure}
\vspace*{20cm}
\hspace*{-3.cm}
\includegraphics{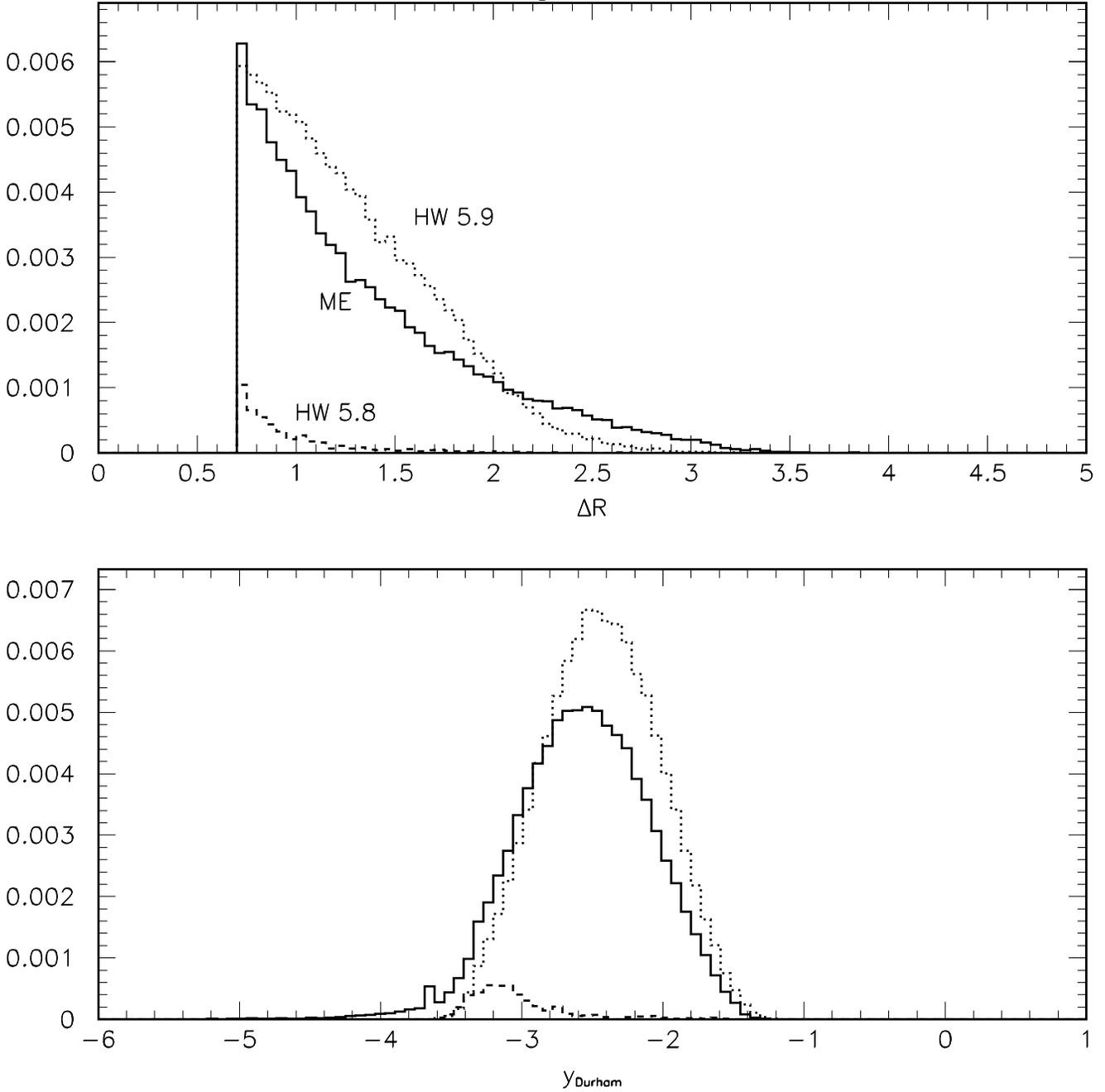}
\vspace{-4cm}
\caption[f7]
{Distributions in (a) the minimum jet-jet angular separation $\Delta R$ and 
(b) the minimum $y$ among jet pairs (defined in the Durham algorithm)
for additional jets produced in top production and decay in
$e^+e^-$ collisions at $\sqrt{s}=360\ {\rm GeV}$. 
Results are shown for the exact calculation (solid histogram, labeled 
ME) and as obtained using HERWIG 
version 5.8 (dashed histogram, labeled HW 5.8) and version 5.9 
(dotted histogram, labeled HW 5.9).
\label{fig:eehw}}
\end{figure}

The results of our comparison for $e^+e^- \rightarrow W^+W^- b \bar b
g$ are shown in Figure \ref{fig:eehw}, where along with the matrix-element
calculation we show results from both the old (5.8) and 
corrected (5.9) versions of HERWIG.  
The center-of-mass energy of 360 GeV is
chosen just above $t \bar t$ threshold to suppress production-stage
emission, so that almost all of the radiation occurs in the decays.
Fig.~\ref{fig:eehw}(a) 
shows the distribution in $\Delta R$ between the closest two
jets, and Fig.~\ref{fig:eehw}(b) shows the minimum $y$ (defined in the Durham
algorithm as $y_{ij} \equiv {{2\min(E_i^2,E_j^2)(1-\cos\theta_{ij})}/
{s}}$) for all jet pairings in the event.  We see in both cases
that the old version of HERWIG underestimates the amount of decay
radiation and the new version overestimates it.  The
discrepancy even in the corrected version of HERWIG is dramatic.

As a technical aside, we note that the normalization of the matrix
element for the $e^+e^-$ case is fixed as in our previous work
\cite{OSSBIS} by choosing the value of $\alpha_s = 0.126$ that gives
agreement between the matrix-element and parton-shower calculations
for the case of $b \bar b$ production. The larger energy scale for top
quark production might suggest the use of a smaller value of
$\alpha_s$, which would make the discrepancy even larger.

The disagreement between the matrix element calculation and HERWIG
seems severe.  A detailed study of the discrepancy is in progress and
will appear elsewhere \cite{OSSINPROG}.  For the moment, it appears
that an estimate of the magnitude of the effect is the best that can
be hoped for.  
We would expect this effect to contribute on the order of a few GeV to 
the uncertainty in the measured top mass.  While not catastrophic, 
clearly such a discrepancy is
unacceptably large, given the precision hoped for in future experiments.
Further work must be done to provide an
accurate event generator for top-quark production.

\section{Conclusions}

Top-quark production will be central to many physics
studies at the LHC, and top mass reconstruction will be the 
key for identifying top events.  
The large energy of the LHC collider provides a
large top-quark cross section, but it also provides for large amounts of
gluon radiation in the top production process.  We have calculated
the cross section for top production and decay in 
association with an extra jet to order $\alpha_s^3$, and find 
a large probability for gluon radiation at the LHC compared to the 
Tevatron.  At the LHC, production-stage radiation 
dominates over decay-stage emissions; this is also in contrast to 
the Tevatron, where the two contributions are roughly comparable.
As shown above, the relative amounts of production- and decay-stage
radiation depend sensitively on the kinematic cuts applied.  In 
addition, the decay contribution is expected roughly to double if 
radiation from hadronic $W$ decays is included.

All of this has important implications for top physics at the LHC.
Even more so than at the Tevatron, gluon radiation at the LHC must be 
understood not only because there is more of it, but because 
uncertainties in quantities like the top mass
will be dominated by systematic effects due to gluon radiation.
For example, the proliferation of production-stage gluon radiation
means that it will sometimes be included in
the top mass reconstruction, and therefore will limit our ability to
reconstruct the top mass.  Quantifying the magnitude of this and similar
effects requires simulations which implement all of the relevant 
physics correctly.  Unfortunately our comparisons show that even the most
recent version of  HERWIG, corrected for the bug in version 5.8,
 still does not reproduce the
correct distributions.  Apparently a hard gluon correction is needed to
model radiation in the production and decay of very heavy quarks.
It should be
a priority to provide a top-quark event generator with the standard
model physics implemented as accurately as possible.

\section*{\Large\bf Acknowledgements}

\noindent WJS is grateful to the UK PPARC for a Senior Fellowship.
Useful discussions with Tony Liss, Richard Partridge and 
Paul Tipton are acknowledged.  This work was supported in part by the U.S.\
Department of Energy, under grant DE-FG02-91ER40685 and by the EU
Programme ``Human Capital and Mobility'', Network ``Physics at High
Energy Colliders'', contract CHRX-CT93-0537 (DG 12 COMA).
\goodbreak

\goodbreak

\newpage

\newpage

\newpage

\newpage

\newpage

\newpage

\newpage

\end{document}